\documentclass[english,aps,prl]{revtex4-2}
\usepackage[T1]{fontenc}
\usepackage[latin1]{inputenc}
\usepackage{graphicx}
\usepackage{amssymb}
\makeatletter
\usepackage{babel}
\makeatother

\usepackage{amsmath}
\usepackage{amsfonts}
\usepackage{tikz}
\usepackage{circuitikz}

\begin{document}
	
\title{What if Planck had known about these calculations?}

\author{Marcos Gil de Oliveira and Kaled Dechoum}

\affiliation{Instituto de F\'{\i}sica - Universidade Federal Fluminense\\
24210-346, Niter\'{o}i-RJ, Brazil}

\begin{abstract}

  Here we present some results that would possibly have attracted the attention of the physics community in the early days of
  quantum mechanics in such a way that its development could have been different from what we see today.
  We will first present a derivation of Planck's blackbody spectrum radiation without the hypothesis of quantized energy levels
  of the oscillator, the only additional hypothesis to the classical theory being the existence
  of zeropoint fluctuations in the electromagnetic field, and this would stand as a nontrivial vacuum which could be inferred by
  classical means.
  After doing this, we derive the Unruh-Hawking effect for the electromagnetic field in a purely undulatory context,
  without dualities and without photons, and one question, among many others, arises after these statements:
  What would be the development of quantum mechanics if Planck was aware of these possibilities?
\end{abstract}
\maketitle

\section{Introduction}

The derivation of the blackbody spectrum due to Planck in 1900 is taken as the birth of a new theory, quantum mechanics,
whose consequences after more than 120 years
are still intriguing, fascinating, and astonishing, besides the fact that its foundations are still a question under discussion.

The consequences of a deep knowledge of the microscopic world have led civilization to many achievements
regarding the understanding of reality and the development of new technologies. However, despite all the knowledge acquired, 
many questions still persist and deserve to be examined. In this article we will insist in one specific issue,
although the reader could think the question is already answered: can quantum phenomena be interpreted by classical means or
do we need to reorganize our logic and our concept of determinism in order to understand the microscopic world?
At least some phenomena can be interpreted by classical means \cite{lapena}, although many others can not.
One of the most impressive examples of quantum phenomena are the entangled states conceived by Schr\"odinger
and easily performed in laboratories today.
These are the typical quantum states whose interpretation has no classical counterpart, according to the vast majority authors,
but reinterpreted by others in the sense of reestablishing the realistic view of quantum mechanics \cite{lapena,kaled1, kaled2}.

A few authors realized that some quantum features of the physical world could be obtained
within the classical framework, simply including in the theory a missing ingredient, the zeropoint field fluctuations,
which could be considered and obtained in classical theory without any reference to the quantum theory \cite{lapena}.
In some sense, these authors believe that quantum mechanics is a consequence of the existence of a nontrivial vacuum, 
but no convincing derivations have been presented so far.

On the other hand, the generation of quantum results employing just classical physics,
empowered by this non trivial vacuum, is a great puzzle and deserves
some attention. In this sense, we propose the question of this article's title,
how would be the development of quantum theory if the perception of the vacuum fluctuations were known in its beginnings?
Quite probably the development would be the same since Planck, in his second theory of the blackbody
spectrum, obtained the missing part in spectrum, due to the vacuum contribution \cite{milonni1}.
However another path could be taken with consequences difficult to imagine, as we shall discuss in the end of this paper.

Most of the present work is based on the ideas of T. H. Boyer on derivation in the spectrum
of blackbody radiation in the realm of stochastic electrodynamics (SED),
a classical electrodynamics theory that includes vacuum fluctuations
as a physical reality \cite{boyer1, boyer2, boyer3, boyer4, boyer5, boyer6},
although we have implemented some of our own ideas. We also follow the achievements presented in the book
of Peter Milloni related to the quantum vacuum in the theory of quantum electrodynamics \cite{milonni1}.
In particular, it is important to quote reference \cite{milonni2}, where we find an elegant and pedagogical derivation of the
Unruh-Davis-Hawking effect, and we follow the same line of reasoning to derive it, but now in the scope
of stochastic electrodynamics, that is, instead of noncommuting bosonic operators describing quantized field,
we use stochastic amplitudes to mimic vacuum fluctuations.

The article follows the sequence: we begin by briefly presenting Planck's achievements and the connections of his work with the
zeropoint field. Then we discuss how the vacuum fluctuation spectrum can be obtained using two fundamental theories,
classical electrodynamics and thermodynamics. This is important since most people think that vacuum fluctuations are strictly an
achievement of quantum mechanics, while we argue here that they can be inferred as a consequence of very well-established
classical theories.

Furthermore, the importance of introducing vacuum fluctuations within the classical framework is that it allows us to derive the
blackbody radiation spectrum without the quantization hypothesis, only assuming that there are two contributions
to the fluctuations of the electromagnetic field in free space, one thermal and the other independent of temperature (vacuum fluctuations),
both uncorrelated (additive), but generating a non-additive energy composition.
This is the main point of the article, since quantum mechanics was introduced by Planck when he quantized the harmonic oscillator's energy
to derive the blackbody radiation spectrum, and here we show another way to obtain the same spectrum, without the quantization assumption.

Finally, we proceed to discuss the vacuum field
in an accelerated frame in order to obtain the Unruh-Davis-Hawking effect in the classical domain,
and we conclude our work with a speculation on how the development of quantum mechanics would have unfolded under this scenario.

\section{A quick glance at Planck's results}

 In this section, we intend to briefly describe Planck's derivation of the blackbody spectrum,
 and point out some connections with vacuum energy. It is interesting to note that, while the blackbody problem
 is celebrated for inaugurating quantum mechanics, it also led to the concept of zeropoint energy.

 To fulfill his purpose of theoretically obtaining the curve that fits the blackbody spectrum,
 Planck \cite{planck} thought of determining the entropy $ S $ of a monochromatic electromagnetic resonator
 as a function of its mean energy  $\left <\mathcal { E} \right>$, that is,
 he was working on what would now be called a microcanonical ensemble.
 Once the relationship was established, one could, from the thermodynamic relation $1/T = dS/d\left< \mathcal{E} \right> $,
 obtain $\left< \mathcal{E} \right>$ as a function of the temperature $T$.

To calculate $S$, Planck made his famous hypothesis that $\left< \mathcal{E} \right>$ must be interpreted as a discrete quantity,
an integer multiple of an energy element $\epsilon$. Using the Boltzmann entropy, $S = k_b \ln W$, where $k_b$ is the Boltzmann constant,
and $W = W(\left< \mathcal{E} \right>)$ is the number of microstates compatible with the system having energy
$\left< \mathcal{E} \right>$, one gets, after a combinatorial calculation,
\begin{equation}
  S = k_b \left[ \left( 1 + \frac{\left< \mathcal{E} \right>}{\epsilon}  \right)
    \ln \left( 1 + \frac{\left< \mathcal{E} \right>}{\epsilon}  \right) -
    \frac{\left< \mathcal{E} \right>}{\epsilon} \ln \frac{\left< \mathcal{E} \right>}{\epsilon}  \right] \,.
\end{equation}

The energy as a function of temperature would then be obtained by calculating
\begin{equation}
  \frac{1}{T} = \frac{d S}{d \left< \mathcal{E} \right> } = \frac{k_b}{\epsilon}
  \ln \left( \frac{\epsilon}{\left< \mathcal{E} \right>} + 1 \right),
\end{equation}
which leads to
\begin{equation}
	\label{marcos1}
	\left< \mathcal{E} \right> =
	\frac{\epsilon}{\exp \left( \epsilon/ k_bT \right) - 1} \,.
\end{equation}

Finally, by applying Wien's law, which states
that $\left< \mathcal{E} \right>(\omega,T) = \omega \, \phi(\omega/T)$ for some single variable function $\phi$,
as we will discuss later, Planck concluded that the energy elements must be proportional to the frequency: $\epsilon = \hbar \omega$.
Briefly, this was the procedure followed by him to derive the blackbody radiation spectrum.

It is expected that, in the limit of high temperatures, one should recover the classical result
of $\left< \mathcal{E} \right> = k_bT$, which is provided by the equipartition theorem. But,
by taking this limit in \eqref{marcos1}, we get
\begin{equation}
  \left< \mathcal{E} \right> \approx \frac{\hbar \omega}{\hbar \omega/ k_bT+\left( \hbar \omega/ k_bT \right)^2/2} =
  \frac{k_bT}{1+\hbar \omega/2 k_bT} \approx k_bT - \frac{\hbar \omega}{2} \,.
\end{equation}
This discrepancy can actually be seen as a first hint of the presence of zeropoint radiation.
If we add to equation \eqref{marcos1} a temperature independent correction of $\hbar \omega / 2$, that is, if we take
\begin{equation}
	\left< \mathcal{E} \right> = \frac{\hbar \omega}{\exp \left( \hbar \omega/ k_bT \right) - 1} + \frac{\hbar \omega}{2} \,,
	\label{marcos2}
\end{equation}
which gives an energy $\left< \mathcal{E} \right>=\hbar \omega / 2$ at $T=0$, the classical result is recovered.

In fact, in 1912, Planck developed a second theory for the blackbody radiation \cite{milonni3},
in which he assumed that the absorption of electromagnetic radiation by a charged oscillator occurs
continuously according to the classical electromagnetic theory, while the 
emission proceeded discontinuously in discrete quanta of energy. Under this hypothesis, when analyzing the balance
in the interaction between radiation and matter, he obtained the expression (\ref{marcos2}) for the energy in each mode
of the spectrum, later confirmed by the quantum electrodynamics.

\section{The electromagnetic vacuum fluctuations}

The reality of the zeropoint fluctuations in the electromagnetic field seems to be an
irrefutable fact. There are many theoretical and experimental evidences for its existence.
Effects such as spontaneous emission, Lamb shift, Casimir effect and some others would be very difficult to
explain without taking into account the electromagnetic fluctuations of the vacuum \cite{milonni1,cohen}.

Regardless of the different assumptions about its origin,
the spectral density of these fluctuations can be deduced in the scope of classical theory,
and also from quantum electrodynamics (QED), where vacuum fluctuations appear as a consequence of the field quantization.
As we show below, both theories yield the same spectral distributions of these fluctuations.

The power spectrum of the radiation field is defined as the product of the
density of modes inside a cavity, that is, the number of modes in the cavity
which supports radiation frequencies between $\omega$ e $\omega + \delta \omega$, by the mean energy that excite
each of these modes. In three dimensions this density of modes scales with $\omega^2$, and the spectral density is given by the expression:
\begin{equation}
\rho ( \omega, T) = \frac{\omega^{2}}{2 \pi^{2} c^{3}} \langle {\cal{E}}\rangle \,.
\label{eq1}
\end{equation}
For instance, if we assume the mean energy being 
$ \langle {\cal{E}}\rangle = k T$ for any normal mode of the field, as prescribed by the energy equipartition
in canonical ensemble, the result will be the Rayleigh-Jeans spectrum.

In the vacuum state ($T =0$) the energy of each mode predicted by quantum theory is $ \hbar \omega / 2 $, and taking into account that each
mode supports two polarization states, the spectrum of the electromagnetic field is:
\begin{equation}
  \rho ( \omega, T=0) = \frac{\hbar \omega^{3}}{2 \pi^{2} c^{3}} \,.
  \label{eq2}
\end{equation}

However, without going beyond thermodynamics and the special theory of relativity,
it is possible to infer the existence of athermic zeropoint fluctuations  of the electromagnetic field given by
equation ({\ref {eq2}}), and this is what we are going to show next.

\subsection{The vacuum spectrum as a Lorentz invariant}

        The zeropoint fluctuation that pervade the whole space as a fundamental background should have the same spectrum for all
        inertial reference frames, since all these frames are equivalent. As a consequence, the electromagnetic vacuum fluctuations
        must be a Lorentz invariant in the following sense: if an observer measures the zeropoint energy density in a frequency interval,
        in a given inertial reference frame, then the observer will find the same spectral energy density in any other inertial frame.
	
	To find the energy spectrum that satisfies this condition, we will follow T.H. Boyer \cite{boyer1, boyer6}.
        Let us write the density of electromagnetic energy of this radiation background through the relation
	\begin{equation}
	  u = \frac{2}{8 \pi} \left< \mathbf{E}^2\right> = \frac{2}{8 \pi} \int d^3 \mathbf{k} \, f(\omega_{\mathbf{k}})
          \label{eq3}
	\end{equation}
	where $u$ is the electromagnetic energy density (energy per unit of volume), $f(\omega_{\mathbf{k}})$ is the energy density
        associated with a given wave vector $\mathbf{k}$, and $\mathbf{E}$ is the electric field vector. 
        The factor $2$ appears due to the fact that, in vacuum, electrical and magnetic energies contribute equally to the total energy.
	
	In another reference system that moves with velocity $\mathbf{v} = v \,\hat{\mathbf{x}}$ along the x-axis in relation to the
        previous one, the components of the four-wave vector are given by
	\begin{eqnarray}
  && \omega_{\mathbf{k}}^\prime =  \gamma (\omega_{\mathbf{k}} - v k_{x}) \, \, \, \, \, \, \, \, \, \, \, \, \,
  k_{x}^\prime =  \gamma (k_{x} - \frac{v}{c^2} \omega_{\mathbf{k}})
  \nonumber \\
  && k_{y}^\prime =  k_{y} \, \, \, \, \, \, \, \, \, \, \, \, \,\, \, \, \, \, \, \, \, \,\, \, \, \, \, \, \, \,\, \, \, \, \, \, \,\,\,
  k_{z}^\prime =  k_{z}  \, ,
  \label{eq5}
        \end{eqnarray}
	while the electromagnetic energy density will be
	\begin{equation}
	  u' = \frac{2}{8 \pi} \left< \mathbf{E}'^2\right> = \frac{2}{8 \pi} \int d^3 \mathbf{k'} f'(\omega'_{\mathbf{k'}})
          \label{eq4}
	\end{equation}

	By Lorentz invariance of the vacuum spectrum we mean that the energy density measured in an narrow interval of frequency is
        the same for all inertial frames, that is $f = f'$. Therefore we should have
	\begin{equation}
	  u' = \frac{2}{8 \pi} \int d^3 \mathbf{k'}\, f(\omega'_{\mathbf{k'}}) = \frac{2}{8 \pi} \int d^3 \mathbf{k'} \,
          f\left[\gamma\left(1-\frac{v k_x}{\omega_\mathbf{k}}\right)\omega_\mathbf{k}\right]
          \label{eq6}
	\end{equation}

	On the other hand, the electric field seen in the new reference frame is related by a Lorentz transformation to the
        electromagnetic field seen in the old frame. This allows us to express $\left< \mathbf{E}'^2\right>$ in terms of quantities
        related to the old frame, as done in \cite{boyer1}, resulting in
	\begin{equation}
	  u' = \frac{2}{8 \pi} \left< \mathbf{E}'^2\right> = \frac{2}{8 \pi}\int d^3\textbf{k}' \,
          f(\omega_\mathbf{k}) \gamma\left(1-\frac{v k_x}{\omega_\mathbf{k}}\right)
          \label{eq4a}
	\end{equation}
	
	Comparing the two expressions for $u'$, equations (\ref{eq6}) and (\ref{eq4a}), we conclude that $f$ must satisfy
	\begin{equation}
	  f\left[\gamma\left(1-\frac{v k_x}{\omega_\mathbf{k}}\right)\omega_\mathbf{k}\right] =
          f(\omega_\mathbf{k}) \gamma\left(1-\frac{v k_x}{\omega_\mathbf{k}}\right) ,
          \label{eq6a}
	\end{equation}
	that is, $f$ must be of the form $f(\omega_{\mathbf{k}}) = \alpha \omega_{\mathbf{k}}$ for some constant $\alpha$.
        We see that this result is exactly the same predicted by quantum mechanics for the harmonic oscillator,
        that is, its energy is proportional to the frequency of the oscillator.
       
        There are, therefore, two invariant spectral densities under Lorentz transformation: one is the trivial vacuum
        with zero energy in each mode, the classical choice, and the other that considers an energy of
        $ \hbar \omega _ {\mathbf {k}} / 2 $ per mode,
        and responsible for many quantum effects. It is important to emphasize that, in this context, the Planck constant is a measure of
        intensity of these fluctuations, and not a quantum of action, as stated by Bohr. 
        We are not saying Bohr was wrong, we are just looking at different contexts of the meaning of Planck's constant.

        With the second choice, the energy density can be written as
        	\begin{equation}
	          u = \frac{2}{8 \pi} \left< \mathbf{E}^2\right> = \frac{2}{8 \pi} \int d^3 \mathbf{k} \, f(\omega_{\mathbf{k}}) =
                  \int_{0}^{\infty} d \omega \, \frac{\hbar \omega^{3}}{2 \pi^{2} c^{3}} \,
          \label{eq6b}
	\end{equation}
                where we have fixed the constant as $\alpha = \hbar/ 2 \pi^{2}$, and due to spatial
                isotropy we write $ d^3 \mathbf{k} = 4\pi k^{2}dk $.
                Finally, we used the plane wave mode relation $kc =\omega_{\mathbf{k}} $.
                The integrand is exactly the zeropoint spectrum, that is invariant under Lorentz transformation.
                
           As can be seen, the vacuum energy diverges when we include the entire frequency spectrum,
           and this is well known in quantum electrodynamics where the renormalization program is used to obtain
           finite physical quantities when matter and electromagnetic field interact, but this discussion is beyond
           the scope of this article.

\subsection{The vacuum spectrum as a consequence of Wien's Law}

Without going beyond thermodynamics and the classical theory of electromagnetism, it is possible to have important
information about the electromagnetic vacuum, based on Wien's displacement law particularized for zero temperature.

Wien's law concerns about the behavior of the radiation spectral density function contained in a bulk,
with perfectly reflecting internal walls, where a quasistatic transformation takes place.
The question to be answered is how the spectral density function
$ \rho (\omega, T) $ changes after an arbitrary quasistatic transformation.
If, for example, the transformation is adiabatic, the change of this function should be given
by the solution of the equation (for details, see reference \cite{landsberg}):
\begin{equation}
\delta \rho (\omega, T) = \left[ \frac{\omega}{3}
 \frac{\partial \rho (\omega, T)}{\partial \omega} - \rho (\omega, T) \right] \frac{\delta V}{V} \,,
 \label{eq7}
\end{equation}
where $V$ is the volume containing the radiation.

This expression must be satisfied by the spectral distribution $ \rho (\omega, T) $ that characterizes electromagnetic radiation at
any temperature $T$ that undergoes an adiabatic transformation. At zero temperature, we must have a ``vacuum'' inside the container.
The vacuum can be described by a spectral density function that does not change ($\delta \rho (\omega, T) =0 $)
even if there is a variation in the volume that contains the radiation.
This implies that there will be no change in the electromagnetic spectral energy density inside the container,
whatever the compression (or expansion) made on the system.

From equation (\ref{eq7}), we see that the spectral distribution at zero temperature must satisfy: 
\begin{equation}
 \frac{\omega}{3}
 \frac{\partial \rho (\omega, T)}{\partial \omega} - \rho (\omega, T) = 0 \,.
 \label{eq8}
\end{equation}

There are two possible solutions for this equation,
$ \rho = 0 $, the vacuum of classical electrodynamics, and
$ \rho = a \omega^{3} $, which is the vacuum of quantum electrodynamics.

The constant $a$ is a proportionality factor associated with the intensity of the field fluctuations
at zero temperature, and to fix this constant to adjust the experimental data associated with the vacuum effect
like Lamb shift and Casimir force, for example, it is useful writing
$ a = \hbar / 2 \pi^{2} c^{3} $, where $ \hbar $
is Planck's constant divided by $ 2 \pi $ and $ c $ is the speed of light in vacuum.

The original (and perhaps better known) form of Wien's law, results that every ``thermodynamically correct''
spectral density function must take the form \cite{lapena}:
\begin{equation}
\rho (\omega, T) = \omega^3 \phi \left(\frac{\omega}{T} \right) \,,
  \label{eq8a}
\end{equation}
where $\phi$ is an arbitrary function of a single variable $\omega/T$.

Although the Wien's law does not determine $\rho (\omega, T)$
(both Planck and Rayleigh-Jeans distribution satisfy it, for example) it is able
to provide us the form of zeropoint spectrum. In fact, assuming that $\lim_{x \to \infty} \phi (x)$ 
is a finite constant $a$, we recover the previous result, $ \rho(\omega,T=0) = a \omega^{3} $.

\section{The Planck spectrum without quantization }

Now we proceed to derive the Planck's blackbody spectrum without the quantization hypothesis.
The starting point is the classical statistical mechanics developed by Boltzmann and Gibbs,
and the existence of athermic fluctuations in the electromagnetic field.

In the canonical ensemble, the system temperature is fixed by a thermal reservoir,
and the system energy fluctuates. The mean thermal energy of the system in contact with a thermal reservoir
at temperature $T$ is given by the expression
\begin{equation}
\langle {\cal{E}}\rangle = \frac{\sum_{r} E_{r} \exp(-\beta E_{r})}{\sum_{r} \exp(-\beta E_{r})} \,,
\label{eq9}
\end{equation}
where $E_{r}$ represents the microscopic energy of a possible state, and the whole energy is distributed
among all possible microstate weighted by the Boltzmann factor, and $\beta = 1/ k_{b}T $.
Thermal fluctuations in energy are, consequently, given by the expression
\begin{equation} 
\left[\langle {{\cal{E}}^{2}}\rangle  - \langle {\cal{E}}\rangle^{2} \right]_{T} = 
- \frac{\partial \langle {\cal{E}} \rangle }{\partial \beta} = k_{b}T^{2}\frac{\partial \langle {\cal{E}}\rangle}{\partial T} \,.
\label{eq10}
\end{equation}

As a hypothesis, we think that the energy fluctuations of each field mode have two origins,
one thermal, dependent on the reservoir temperature, 
and other athermic, universal and independent of temperature.
These fluctuations are assumed to be statistically independent, so that we can write:
\begin{equation}
\left[\langle {{\cal{E}}^{2}}\rangle - \langle {\cal{E}}\rangle^{2} \right]_{T} =
\left[\langle {{\cal{E}}^{2}}\rangle - \langle {\cal{E}}\rangle^{2} \right]_{Tot} -
\left[\langle {{\cal{E}}^{2}}\rangle - \langle {\cal{E}}\rangle^{2} \right]_{ZP} \,,
\label{eq11}
\end{equation}
where these three terms represent the thermal, total and zeropoint fluctuations, respectively.

At this point it is worth remembering the principle of maximum entropy, which states that in possession of the incomplete
information we have about the system, where we only know that it
is in thermodynamic equilibrium, the best probability distribution we can adopt is such that $ P({\cal{E}}) $
maximizes the functional \cite{jaynes}:
\begin{equation}
    S[P(\mathcal{E})] = -k_b \int_{0}^{\infty} P(\mathcal{E}) \ln \left[P(\mathcal{E}) \right] d \mathcal{E} \,,
 \label{eq12}
\end{equation}
with the constraints that the distribution should be properly normalized, and that the expected value for the energy
$\left< \mathcal{E} \right>$ is fixed, a condition that is enforced by the contact with the thermal reservoir.
Therefore, it is possible to show that $P(\mathcal{E})$ must be given by 
\begin{equation}
  P(\mathcal{E}) = \frac{1}{\left< \mathcal{E} \right>}e^{-\mathcal{E}/\left< \mathcal{E} \right>} \,.
\label{eq13}
\end{equation}

Using equation (\ref{eq13}) for the probability distribution, we obtain
\begin{equation}
  \left< \mathcal{E}^2 \right> = \int_0^\infty \mathcal{E}^2 P(\mathcal{E}) d \mathcal{E} = 2
  \left< \mathcal{E} \right>^2 \,,
  \label{eq14}
\end{equation}
and, therefore, the variance results in
$\sigma^2 = \left< \mathcal{E}^2 \right> - \left< \mathcal{E} \right>^2 = \left< \mathcal{E} \right>^2$.

Thus, using the definition of spectral density function, equation (\ref{eq1}), the equation (\ref{eq2}) for the vacuum spectrum,
equation (\ref{eq14}), and the expression for thermal fluctuations in the canonical  ensemble, equation (\ref{eq10}),
we can write a differential equation for the spectral density function $\rho$ as follows,
\begin{equation}
k T^{2} \frac {\partial \rho}{\partial T} = \frac{\pi^{2} c^{3}}{\omega^{2}} \left[\rho^{2} 
- \left( \frac{\hbar \omega^{3}}{2 \pi^{2} c^{3}}\right)^{2}\right] \,.
\label{eq15}
\end{equation}

If we take $\hbar \rightarrow 0$, the solution of equation (\ref{eq15}) is the
Rayleigh-Jeans spectrum, however, if we take the scale parameter of vacuum fluctuations $\hbar$
to be given by Planck's constant, the solution will be given by the following spectral function
(in both cases, the solution is given by the method of separation of variables)
\begin{equation}
\rho (\omega, T) = \frac{\hbar \omega^{3}}{2 \pi^{2} c^{3}} \coth \left(\frac{\hbar \omega}{2 k_b T} \right) 
= \frac{\hbar \omega^{3}}{ \pi^{2} c^{3}} \left[ \frac{1}{2} + \frac{1}{\exp{(\frac{\hbar \omega}{k_bT})}-1} \right] \,.
\label{eq16}
\end{equation}

This is exactly the blackbody radiation spectrum with addition of the zeropoint vacuum spectrum, that correspond to a mean energy
\begin{equation}
\langle {\cal{E}}\rangle =  \frac{\hbar \omega}{2} + \frac{\hbar \omega}{\exp{(\frac{\hbar \omega}{k_bT})}-1} \,,
\label{eq16a}
\end{equation}
for each normal mode.

It is interesting to note that the blackbody radiation spectrum was obtained assuming that the field has
two statistically independent sources of energy fluctuations, while the average energy of each electromagnetic
mode appears as a combination, not a simple sum.
This is because, in this case, you cannot satisfy both the additivity of the variances and the additivity
of the energies for the same system. The physical meaning of this, however, needs further study.

\section{The vacuum seen from a constant accelerated reference frame}

One of the most striking effects associated with the existence of a nontrivial vacuum is its behavior seen from a
reference frame in accelerated motion or at rest in a gravitational field.
As we have seen, the vacuum spectrum is a Lorentz invariant  quantity, any observer in an inertial reference frame
sees the same vacuum spectrum. However, when viewed from a uniformly accelerated reference frame,
the new vacuum spectrum seems to come from a thermal source.
This result was obtained by Unruh and Davis, in the context of quantum field theory, and by Hawking in the context of
black hole evaporation \cite{milonni1}, and suggests the appearance of photons,
which are quantum excitation of the electromagnetic field (and are particles in the quantum sense),
due to the state of motion of the frame, and this is very intriguing.
 
Here we propose to obtain the same result in the scope of classical field theory.
For this purpose we will follow the same construction done in reference \cite{milonni2},
in a very nice and pedagogical derivation, but now using stochastic c-number field amplitudes instead of field operators,
and pointing out the differences between the two interpretations. 

\subsection{Motion with an uniform relativistic acceleration}

The uniformly accelerated relativistic motion is one in which the acceleration $a$ is constant, 
and the four-acceleration in the proper reference frame is given by
\begin{equation}
a^{\mu} = (0,a,0,0) \,.
\end{equation}
Using the Lorentz transformation law for four-vectors 
(let's imagine an inertial frame that will be instantly at rest in
relation to the accelerated frame), we have
\begin{equation}
a^{\nu} = {\cal L}^{\nu \mu} a_{\mu} = (\beta \gamma a,  \gamma a, 0 , 0) \,.
\label{eq17}
\end{equation}
where $ {\cal L}^{\nu \mu} $ is the $4 \times 4$ Lorentz transformation matrix, $\beta = v/c$, and $\gamma = (1 - v^{2}/c^{2})^{-1/2}$.
Note that  $a^{\mu}a_{\mu} = -a^{2}$ is an invariant (scalar) quantity.

In an inertial frame, the movement of a particle with constant proper acceleration
should be described by equation
\begin{equation}
a^{\mu} = \frac{d v^{\mu}}{d \tau} = \left( \frac{d}{d \tau} \left[ \frac{c}{\sqrt{1 - v^2/c^2}} \right] \,, 
\frac{d}{d \tau}\left[ \frac{\vec{v}}{\sqrt{1 - v^2/c^2}} \right] \right) \,.
\label{eq18}
\end{equation}

From equations (\ref{eq17}) and (\ref{eq18}) one can write, assuming motion in the $x$ direction, that is, $v = v_{x}$,
the following equation,
\begin{equation}
\gamma a =  \gamma \frac{d}{d t}\left[ \frac{v}{\sqrt{1 - v^2/c^2}} \right] \,.
\label{eq19}
\end{equation}

This equation can be integrated assuming as initial condition
that the particle was at rest when the movement started, and we obtain
\begin{equation}
  v(t) = \frac{at}{\sqrt{1 + a^2 t^2/c^2}} \,,
  \label{eq20}
\end{equation}
and integrating once more, we obtain the position function, 
\begin{equation}
x(t) = \frac{c^2}{a} \left( \sqrt{1 + a^2 t^2/c^2} - 1 \right) \,.
\label{eq21}
\end{equation}

We can express the last expressions in terms of the proper time  $\tau$
using the relation
\begin{equation}
  \frac{dt}{d \tau} = \gamma \,\,\,\,\, \Rightarrow  \,\,\,\,\, \tau = \int_{0}^{t} \sqrt{1 - v^2/c^2} dt^\prime  \,\,\,\,\,
  \Rightarrow   \,\,\,\,\,
t (\tau) = \frac{c}{a} \sinh \left(\frac{a\tau}{c} \right) \,,
\label{eq22}
\end{equation}
where in the second step we used equation (\ref{eq20}),
and therefore the functions of the velocity and position parameterized by the proper time are given by
\begin{eqnarray}
 v(\tau) &=& c \tanh \left( \frac{a \tau}{c} \right) \nonumber \,, \\
 x(\tau) &=& \frac{c^2}{a} \left[ \cosh \left( \frac{a \tau}{c} \right) - 1 \right] \,.
 \label{eq23}
\end{eqnarray}

These functions characterize the relativistic motion of a constant proper aceleration, and as expected the velocity 
tends asymptotically to $c$, speed of light, and the position grows exponentially with the proper time.

\subsection{A plane wave seen from an accelerated frame}

Before going on to the problem itself, that is, determining how the
the vacuum spectrum is seen from an accelerated frame,
let's ask how a monochromatic plane wave, given by the expression
\begin{equation}
\vec{E} (x,t) = E_{0} e^{i(kx-\omega t)} \hat{e}_{y} \,,
\label{eq24}
\end{equation}
is seen in an accelerated reference frame.

It is known that when we change from one inertial reference frame to another,
a plane wave remains a plane wave, and the angular frequency of this wave,
as seen before, transforms as follows
\begin{equation}
\omega^{\prime} = \gamma \omega \left( 1 - \frac{k v}{\omega}\right) \,.
\label{eq25}
\end{equation}

The expression (\ref{eq25}) shows how the frequency of a monochromatic wave
is transformed from one inertial frame to another, which is called the
Doppler effect. However, we can go further and imagine that, in the case of an accelerated referential, 
there will always be an inertial reference frame that instantly has the
same speed of the latter in such a way that the Doppler effect changes
continuously in the time parameterized by $\tau$ \cite{milonni2, moriconi},
\begin{equation}
\omega^{\prime} (\tau) = \gamma (\tau) \omega \left( 1 - \frac{k v(\tau)}{\omega}\right) \,.
\label{eq26}
\end{equation}
where $ \omega / k = \pm c $, depending on whether the plane wave moves in the same direction or opposite to the observer.
Using the velocity function for the uniformly accelerated motion
obtained before, equation (\ref{eq23}), we get 
\begin{equation}
\omega^{\prime} (\tau) = \omega \frac{1 - \tanh \left( \frac{a \tau}{c} \right) }{\sqrt{1 - \tanh^{2} \left( \frac{a \tau}{c} \right)}} = 
\omega e^{-a \tau/c} \,,
\label{eq27}
\end{equation}
while in case of the frame moving in the opposite direction to the wave we have $\omega^{\prime} (\tau) =\omega e^{+a \tau/c} $.
Expressions (\ref{eq26}) and (\ref{eq27}) are referred to a time-dependent Doppler shift observed in an accelerated frame.
In summary, when the observer is moving in the same direction as the plane wave, he observes an exponential decrease in its frequency,
while when traveling in the opposite direction, the frequency increases exponentially.

\subsection{The vacuum spectrum seen from an accelerated frame}

In order to obtain the spectrum of the vacuum seen from an accelerated reference frame, we will make use of the Wiener-Kintchine
theorem \cite{pathria, mandel} which says that the power spectrum of the radiation field (for a stochastic process, in general)
is obtained from the Fourier transform of the time correlation function of the field.

In order to state more precisely this theorem, consider a stationary stochastic process denoted by $ g (t) $,
which fluctuates in time. The time correlation function is defined as
\begin{equation}
\Gamma (\tau) = \lim_{T \rightarrow \infty} \frac{1}{2T} \int_{- T}^{T} g^{*}(t) g(t + \tau) \, dt \, .
\end{equation}
Note that, being a stationary process, $\Gamma$ does not depend on a specific instant,
but only on the time interval $ \tau $ between two events.

The power spectrum  $S (\Omega)$ associated with a stochastic process $ g (t) $ is then defined as
\begin{equation}
  S(\Omega) = \frac{1}{2 \pi} \int_{-\infty}^{\infty} \Gamma (\tau) e^{-i \Omega \tau} \, d\tau \,,
\end{equation}
and, in this sense, $S(\Omega)$ will measure the intensity fluctuation associated with a Fourier component $\Omega$ of $g(t)$.

As we are dealing with the stationary process, we can change from an average over time, after a very long time process,
to an average over an ensemble of all possible realizations, where we fix two instants and perform the average
over the whole ensemble. In this way, the time correlation can be obtained from the relation
\begin{equation}
\Gamma (\tau) =  \langle g^{*}(t) g(t + \tau) \rangle \, ,
\end{equation}
where the brackets mean ensemble average. As a consequence, the power spectrum of the fluctuations is obtained from the relation
\begin{equation}
  S(\Omega) \delta (\Omega - \Omega^\prime) =  \langle g^*(\Omega) g(\Omega^\prime)  \rangle \,,
\end{equation}
where $g(\Omega)$ is the Fourier transform of $g(t)$.

There is a slight difference between power spectrum $ S(\Omega)$ defined here and spectral density $\rho (\Omega)$
defined before in equation (\ref{eq1}) , and without going too far, here we just put the
conversion \cite{mandel}: $  S(\Omega) = 2 \pi/3 \, \rho (\Omega) $.

Now we turn our attention to the fluctuating field that permeates the whole space at zero Kelvin, the zeropoint field.
To avoid confusion with cumbersome vector calculations, let's deal with a scalar field in one dimension described by
\begin{equation}
g (t) = \sum_{\mathbf{k}} 
\left(  \frac{2 \pi \hbar c^2}{\omega{_\mathbf{k}} V } \right)^{1/2} \left[ \alpha_{_{\mathbf{k}}} e^{-i \omega_{\mathbf{k}}t} \ + 
\alpha^{*}_{_{\mathbf{k}}} e^{i \omega_{\mathbf{k}}t} \right] \,,
\label{eq28}
\end{equation}
where the sum is over all possible wavevectors, and $\alpha_{_{\mathbf{k}}}$ is the complex amplitude
of the mode  $\mathbf{k}$ whose phase is assumed to be randomly distributed in the interval $(0, 2 \pi) $, satisfying the relation
\begin{equation}
  \langle \alpha^{*}_{_{\mathbf{k}}}\alpha_{_{\mathbf{k}^{\prime}}} \rangle = \frac{1}{2}   \delta_{\mathbf{k}, \mathbf{k}^{\prime} } \,,
\label{eq29}
\end{equation}
the average being taken over the ensemble, and any other correlation is zero. We are working with
the field in the vacuum state, and that is why only the term $1/2$ will be present in equation (\ref{eq29}).
Otherwise we should include the thermal part.

The Fourier transform of $ g (t) $ will be given by
\begin{eqnarray}
g(\Omega) &=& \frac{1}{2 \pi} \int^{\infty}_{-\infty} dt\,g (t) e^{-i \Omega t} \nonumber \\
&=& \frac{1}{2 \pi} \int^{\infty}_{-\infty} dt\,\sum_{\mathbf{k}} 
\left( 
\frac{2 \pi \hbar c^2}{\omega{_\mathbf{k}} V } \right)^{1/2} \left[ \alpha_{_{\mathbf{k}}} e^{-i \omega_{\mathbf{k}}t} \ + 
\alpha^{*}_{_{\mathbf{k}}} e^{i \omega_{\mathbf{k}}t} \right]\, e^{-i \Omega t} \,.
\label{eq30}
\end{eqnarray}

We will now consider an observer with an uniform proper acceleration in the
the vacuum field. This observer measures each frequency of the field shifted by a Doppler effect, and for this observer
$ g (\Omega) $ has the form
\begin{eqnarray}
  g(\Omega) =  \frac{1}{2 \pi}
\int^{\infty}_{-\infty} d\tau\, \sum_{\mathbf{k}} 
\left( 
\frac{2 \pi \hbar c^2}{\omega{_\mathbf{k}} V }  \right)^{1/2}
\left[ \alpha_{_{\mathbf{k}}} e^{-i \int^{\tau} \omega_{\mathbf{k}}(t)\, dt} \, + \,
\alpha^{*}_{_{\mathbf{k}}} e^{i \int^{\tau} \omega_{\mathbf{k}} (t) \,dt} \right]\, e^{-i \Omega \tau} \,,
\label{eq31}
\end{eqnarray}
where the phase $\varphi(\tau) =\int^{\tau} \omega_{\mathbf{k}}(t)\, dt $ is integrated due to the constant
change of the angular frequency with respect the proper time.

Using equation (\ref{eq27}), the expression for $g (\Omega)$ is then written as
\begin{eqnarray}
  g(\Omega) =  \frac{1}{2 \pi} 
  \sum_{\mathbf{k}} 
\left( 
\frac{2 \pi \hbar c^2}{\omega{_\mathbf{k}} V } \right)^{1/2}
 \int^{\infty}_{-\infty} d\tau\,
\left[ \alpha_{_{\mathbf{k}}} 
e^{i ( \omega_{\mathbf{k}} c/a) e^{-a \tau/c}} \, 
+ \,
\alpha^{*}_{_{\mathbf{k}}} e^{-i( \omega_{\mathbf{k}} c/a) e^{-a \tau/c }} \right]\, e^{-i \Omega \tau} \,.
\label{eq32}
\end{eqnarray}

In order to integrate equation (\ref{eq32}), we consider the new variable $y = \omega_{\mathbf{k}}ce^{-a\tau/c}/a$,
in terms of which we have
\begin{equation}
  \int^{\infty}_{-\infty} d\tau\,e^{i \Omega \tau}\,e^{\pm i( \omega_{\mathbf{k}} c/a) e^{-a \tau/c }} = \frac{c}{a}
  \left( \frac{a}{\omega_{\mathbf{k}} c} \right)^{i \Omega c/a}\int_{0}^{\infty}y^{i \Omega c/a - 1} e^{\pm i y} dy .
\end{equation}

The calculation of this integral is lengthy, and details can be found in the appendix. The result is
\begin{equation}
	\int_{0}^{\infty}y^{p - 1} e^{- i y} dy = (-i)^p\Gamma(p),
\end{equation}
where $\Gamma$ is the Gamma function.

With this formula we can go back and write the expression for the scalar field in one dimension seen by the
uniform accelerated frame, which takes the following form:
\begin{equation}
g(\Omega) =  \frac{1}{2 \pi}\sum_{\mathbf{k}} \left( \frac{2 \pi \hbar c^2}{\omega{_\mathbf{k}} V } \right)^{1/2}\frac{c}{a}
\Gamma \left(i \frac{ \Omega c}{a} \right) 
 \left[ e^{- \pi \Omega c/2a} \alpha_{_{\mathbf{k}}} + e^{\pi \Omega c/2a} \alpha^{*}_{_{\mathbf{k}}} \right] 
\left(\frac{a}{\omega_{\mathbf{k}} c} \right)^{i \Omega c/a} \,.
\label{eq34}
\end{equation}

In possession of $g(\Omega)$, we are now in a position to obtain the spectral density of the field.
The Fourier transform of the time correlation function of the present scalar field is then given by
\begin{eqnarray}
 \langle g(\Omega)  g^{*}\left(\Omega^{\prime}\right) \rangle &=&  \left( \frac{c}{2 \pi a} \right)^{2}    
\Gamma \left(i \frac{ \Omega c}{a} \right)  \Gamma \left(-i \frac{ \Omega^{\prime} c}{a} \right)
\nonumber \\
& \times & \sum_{\mathbf{k}} \sum_{\mathbf{k}^{\prime}} \left( \frac{2 \pi \hbar c^2}{ V } \right) 
\frac{1}{\sqrt{\omega_{\mathbf{k}}\omega_{\mathbf{k}^{\prime}}}} \left(\frac{a}{\omega_{\mathbf{k}} c} \right)^{i \frac{\Omega c}{a}}
\left(\frac{a}{\omega_{\mathbf{k}^{\prime}} c} \right)^{-i \frac{\Omega^{\prime} c}{a}} \nonumber \\
&\times & \left<  \left[ e^{ -\pi \Omega c/2a} \alpha_{_{\mathbf{k}}} + e^{ \pi \Omega c/2a} \alpha^{*}_{_{\mathbf{k}}} \right] 
 \left[ e^{ -\pi \Omega^{\prime} c/2a} \alpha^{*}_{_{\mathbf{k}^{\prime}}} + e^{ \pi \Omega^{\prime} c/2a} \alpha_{_{\mathbf{k}^{\prime}}} \right] 
 \right> \,.
 \label{eq35}
\end{eqnarray}

The brackets act on the random amplitudes, and equation (\ref{eq29}) tell us that different wavevectors are uncorrelated,
allowing us to eliminate one of the summations in the equation (\ref{eq35}),
\begin{eqnarray}
 \langle g(\Omega)  g^{*}\left(\Omega^{\prime}\right) \rangle &=&  \left( \frac{c}{2 \pi a} \right)^{2}    
\Gamma \left( i \frac{ \Omega c}{a} \right)  \Gamma \left(-i \frac{ \Omega^{\prime} c}{a} \right)
\nonumber \\
& \times & \sum_{\mathbf{k}} \left( \frac{2 \pi \hbar c^2}{ V } \right) 
\frac{1}{\omega_{\mathbf{k}}} \left(\frac{a}{\omega_{\mathbf{k}} c} \right)^{i \frac{\Omega c}{a}}
\left(\frac{a}{\omega_{\mathbf{k}} c} \right)^{-i \frac{\Omega^{\prime} c}{a}} \nonumber \\
&\times &  \frac{1}{2} \left[ e^{ \pi (\Omega + \Omega^{\prime} ) c/2a} + e^{ - \pi  (\Omega + \Omega^{\prime} ) c/2a} \right] \,.
\label{eq36}
\end{eqnarray}

At this point it is convenient perform the limit to continuum by transforming the sum over
wavevectors into an integral following the relation
$$ \frac{1}{V} \sum_{\mathbf{k}} = \left ( \frac{1}{2 \pi} \right )^{D} \int d^{D} {\mathbf{k}}  $$
where $D$ represents the dimension of the system.

In the simplest case we are working on, $D = 1$, we can easily evaluate the integral
\begin{equation}
  I = \frac{1}{2 \pi c} \int_{0}^{\infty} d \omega_{\mathbf{k}} \, \frac{1}{\omega_{\mathbf{k}}}
  \left(\frac{a}{\omega_{\mathbf{k}} c} \right)^{i \frac{\Omega c}{a}}
\left(\frac{a}{\omega_{\mathbf{k}} c} \right)^{-i \frac{\Omega^{\prime} c}{a}} \,,
\label{eq37}
\end{equation}
where we have used  $ \omega_{\mathbf{k}}  =kc $.
Performing the change of variable $ x = \ln (\omega_{\mathbf{k}} c/a) $, the integral can be calculated resulting in
\begin{equation}
\int_{-\infty}^{\infty} dx \, e^{-ix(\Omega - \Omega')c/a} = 2 \pi \frac{a}{c} \delta (\Omega - \Omega') \,.
\label{eq38}
\end{equation}

Thus, the correlation function will be given by
\begin{eqnarray}
 \langle g(\Omega)  g^{*}\left(\Omega^{\prime}\right) \rangle &=&  \left( \frac{c}{2 \pi a} \right)^{2} 2 \pi \hbar a   
 \left | \Gamma \left(i \frac{ \Omega c}{a} \right) \right | ^{2}  \cosh \left( { \frac{\pi \Omega c}{a}} \right) \,
 \delta (\Omega - \Omega') \,.
\label{eq39}
\end{eqnarray}

Using the following relation for gamma function with an imaginary argument \cite{arfken}:
\begin{equation}
\left | \Gamma \left(i \frac{ \Omega c}{a} \right) \right |^{2} = \frac{\pi}{(\Omega c/ a) \sinh (\pi \Omega c/ a)} \,,
\label{eq40}
\end{equation}
we finally get the result
\begin{equation}
\langle g^{*}\left(\Omega^{\prime}\right)g(\Omega)\rangle = \frac{ \hbar c}{2 \Omega} \coth \left( { \frac{\pi \Omega c}{a}} \right)
\delta (\Omega - \Omega^{\prime}) \equiv S(\Omega) \, \delta (\Omega - \Omega^{\prime}) \,.
\label{eq41}
\end{equation}

This spectrum is the same as the thermal blackbody spectrum 
with inclusion of the zeropoint term, see equation (\ref{eq16}),
and we recognize the ``temperature'' 
\begin{equation}
T = \frac{\hbar a}{ 2 \pi k_{b} c} \,.
\end{equation}
The difference between the pre-factor in the equation (\ref{eq16}) and the equation (\ref{eq41}) is related to dimensionality,
whereas here we are working in one dimension, there we were working in three dimensions.

What we have just seen here is that although the vacuum spectrum is a Lorentz invariant, when viewed from a constantly
accelerated frame of reference (or at some point in a gravitational field, as the equivalence principle does not distinguish
these two cases) a thermal spectrum is perceived, where instead of the temperature of the space we have
an effective temperature proportional to the proper acceleration.

This phenomenon when associated with intense gravitational fields is known as black hole evaporation,
and in field theory as the Unruh-Davis effect, and is another surprising effect due to the existence of a nontrivial vacuum.

\section{Some comments and Conclusions}

What we have seen in this article is that quantum mechanics could have a different starting point if the existence of  vacuum fluctuations
was predicted at the beginning, as suggested by Nernst, Planck, Einstein and others around the 1910s \cite{milonni1}.
As Planck said, postulating the quantization of energy was an ``act of desperation'', it worked, but the reason for it has not
been satisfactorily explained, and it may be asked whether he would have preserved the continuous energy and maintained the
entire classical theory with just the addition of a nontrivial vacuum.

We have also shown that the existence of a nontrivial vacuum spectrum can be inferred in at least two distinct ways,
either by requiring the Lorentz invariance of the vacuum spectrum or by the Wien displacement law based on thermodynamics.
Both predict the same two possibilities: (i) the classical vacuum, which is ``nothing'' or (ii) a nontrivial vacuum,
the same present in the usual quantum theory, full of fluctuations and responsible for the many effects
observed today in laboratories around the world. 

Planck's blackbody radiation spectrum could be derived without using the quantization hypothesis of the oscillator
energy levels, the existence of a nontrivial vacuum spectrum that permeates all space,
being Lorentz invariant, was enough.
However, a connection between the existence of a nontrivial
vacuum and the quantization of the energy is a point to be questioned and researched.

The Unruh-Davis-Hawking effect was obtained in a purely classical wave theory, without any dualistic context.
Here, what is a wave in one frame is still a wave in the other, and no particles were needed in theory,
while in the quantum mechanics view, photons, which are seen as particles that excite the field, appear due to the change
from an inertial frame to a non-inertial frame. We are not defending either theory, but just pointing out some
strangeness of a particle concept that appears when the reference frame is changed from inertial to accelerated.

Another important issue to be highlighted is that any field has its own vacuum \cite{aitchinson},
and although we are talking about electromagnetic field, some of these ideas could be thought for other fields.

All these results may suggest that quantum mechanics is a consequence of the existence of a nontrivial vacuum.
Although this statement is very strong, it is not unrealistic and there are some scientists who support it
(see \cite{dice} and references therein), but always having in mind that
quantum mechanics is a successful theory, that still seeks a more solid foundation. Finally, a natural question can be asked:
If Planck knew about these results, would the development of quantum mechanics be any different?

\section*{Acknowledgments}

We thank Reinaldo Faria De Melo E Souza and Marco Moriconi for fruitful discussions and critical reading of the manuscript,
and Instituto Nacional de Ci\^encia e Tecnologia de Informa\c c\~ao Qu\^antica (INCT-CNPq).

\subsection*{Appendix}

 In this appendix, we intend to show that
\begin{equation}
	\label{complex_integral_for_gamma}
	\int_{0}^{\infty} t^{p-1}e^{-it} dt = (-i)^p \Gamma(p) \ \ \text{ for } \Re(p) \in (0,1),
\end{equation}
where and $\Gamma$ is the Gamma function. Although the condition over $\Re(p)$ is imposed to ensure convergence,
the right hand side of \eqref{complex_integral_for_gamma} is perfectly well behaved if $p$ is a purely imaginary number,
that is, if $\Re(p) = 0$ and $p \ne 0$. Therefore, even in this case, we can regularize the integral by taking its
value to be $(-i)^p \Gamma(p)$, which is the result used in the text.

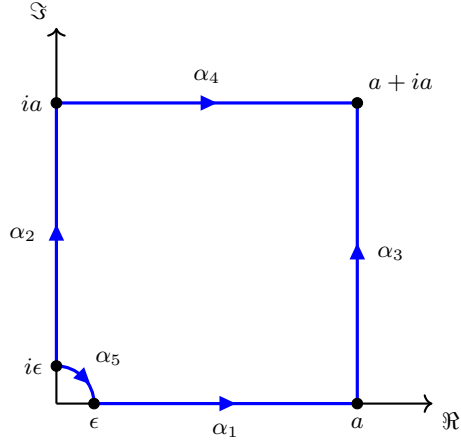
\begin{figure}
	\centering
	\begin{tikzpicture}
		
		\draw[->,thick] (0,0) -- (5,0) node[anchor=north west] {$\Re$};
		\draw[->,thick] (0,0) -- (0,5) node[anchor=south east] {$\Im$};

		\draw[blue,very thick] (0.5,0) -- (4,0)
		node[color=blue,
		currarrow,
		pos=0.5, 
		xscale=1,
		sloped,
		scale=1] {}
		node[color = black,below=4pt,pos=0.5] {$\alpha_1$};
		
		\draw[blue,very thick]  (4,0) -- (4,4)
		node[color=blue,
		currarrow,
		pos=0.5, 
		xscale=1,
		sloped,
		scale=1] {}
		node[color = black,right=4pt,pos=0.5] {$\alpha_3$};
		
		\draw[blue,very thick] (4,4) -- (0,4)
		node[color=blue,
		currarrow,
		pos=0.5, 
		xscale=1,
		sloped,
		scale=1] {}
		node[color = black,above=4pt,pos=0.5] {$\alpha_4$};
		
		\draw[blue,very thick] (0,0.5) -- (0,4)
		node[color=blue,
		currarrow,
		pos=0.5, 
		xscale=1,
		sloped,
		scale=1] {}
		node[color = black,left=4pt,pos=0.5] {$\alpha_2$};
		
		\draw[blue,very thick] (0,0.5) arc (90:0:0.5cm)
		node[color=blue,
		currarrow,
		pos=0.5, 
		xscale=1,
		sloped,
		scale=1] {}
		node[color = black,above right=1pt,pos=0.5] {$\alpha_5$};

		\filldraw[black] (4,0) circle (2pt) node[below=2pt] {$a$};
		\filldraw[black] (0.5,0) circle (2pt) node[below=2pt] {$\epsilon$};
		\filldraw[black] (0,0.5) circle (2pt) node[left=2pt] {$i\epsilon$};
		\filldraw[black] (0,4) circle (2pt) node[left=2pt] {$ia$};
		\filldraw[black] (4,4) circle (2pt) node[above right=1pt] {$a+ia$};
	\end{tikzpicture}
	\caption{Contour described by equation \eqref{paths_eq}.}
	\label{paths_image}
\end{figure}

In order to obtain \eqref{complex_integral_for_gamma}, we first consider the curves in $\mathbb{C}$ shown
in figure \ref{paths_image}, which are defined by the functions
\begin{equation}
	\label{paths_eq}
	\begin{matrix}
		\alpha_1(t) &=& t; & \,\,\,\,\,\,\,  \alpha_2(t) &=& i t; \\
		\alpha_3(s) &=& a+i s; &  \,\,\,\,\,\,\, \alpha_4(s) &=& s + i a; \\
		\alpha_5(\varphi) &=& i \epsilon e^{-i \varphi},
	\end{matrix}
\end{equation}
where $a,\epsilon$ are positive real numbers, $t\in [\epsilon,a], \ s \in [0,a]$ and $\varphi \in [0,\pi/2]$.
Now, let $f(z) = z^{p-1} e^{-z}$. As $f$ is holomorphic in $\mathbb{C} \setminus \{0\}$, we obtain, by Cauchy's integral theorem,
\begin{equation}
	\label{cauchy}
	\sum_{j=1}^{5} (-1)^j I_j = 0,
\end{equation}
where
\begin{equation}
	I_j = \int_{\alpha_j} f(z) dz.
\end{equation}

Notice that, in the limit $a \to \infty$ and $\epsilon \to 0$, we have
\begin{equation}
	I_1 = \int_{0}^{\infty} t^{p-1} e^{-t} dt,
\end{equation}

which is the definition of $\Gamma(p)$ when $\Re(p)>0$. On the other hand,
\begin{equation}
  I_2= i^p \int_{0}^{\infty} t^{p-1} e^{-it} dt.
  \label{integral}
\end{equation}
Therefore, we can from \eqref{cauchy} conclude \eqref{complex_integral_for_gamma},
provided that we are able to show that $I_3,I_4$ and $I_5$ tend to zero in the considered limit, an issue that we will address now.

All of the integrals will be treated in the same way. First we make use of the property that
\begin{equation}
	\left| \int_a^b F(t) dt \right| \le \int_a^b \left| F(t) \right|dt
\end{equation}
for any continuous function $F:[a,b] \to \mathbb{C}$.
We proceed by estimating the integral represented in equation (\ref{integral}) to show that the estimate tends to zero.

Let $x = \Re(p)$ and $y = \Im (p)$.

\begin{itemize}
	\item $\mathbf{I_3}$
	
	We write $\alpha_3(s) = \sqrt{a^2 + s^2}e^{i \theta}$, with $\theta = \arctan(s/a) \in [0,\pi/4]$. Then, we have
	\begin{equation}
		I_3 = i \int_0^a \left(\sqrt{a^2+s^2}\right)^{p-1}e^{i \theta (p-1)} e^{-(a+is)} ds,
	\end{equation}
	and, therefore,
	\begin{equation}
		\begin{aligned}
			\left | I_3 \right | &\le \int_0^a \left(\sqrt{a^2+t^2}\right)^{x-1}e^{- \theta y} e^{-a} dt\\
			&\le a \left( \sqrt{2} a \right)^{x-1} e^{\pi |y| / 4} e^{-a},
		\end{aligned}
	\end{equation}
	which tends to zero when $a \to \infty$.
	
	\item $\mathbf{I_4}$
	
	We write $\alpha_4(t) = \sqrt{a^2 + s^2}e^{i \theta}$ with $\theta = \arctan(a/s) \in [\pi/4,\pi/2]$. This time, we have
	\begin{equation}
		I_4 = \int_0^a \left(\sqrt{a^2+s^2}\right)^{p-1}e^{i \theta (p-1)} e^{-(s+ia)} ds,
	\end{equation}
	and, then,
	\begin{equation}
		\begin{aligned}
			\left | I_4 \right | &\le \int_0^a \left(\sqrt{a^2+s^2}\right)^{x-1}e^{- \theta y} e^{-s} ds\\
			&\le \left( \sqrt{2} a \right)^{x-1} e^{\pi |y| / 2}\int_0^ae^{-s} ds\\
			&= \left( \sqrt{2} a \right)^{x-1} e^{\pi |y| / 2} \left(1-e^{-a}\right),
		\end{aligned}
	\end{equation}
	which, as $x <  1$, tends to zero when $a \to \infty$.
	
	\item $\mathbf{I_5}$
	
	We have
	\begin{equation}
		I_5 = - i (i\epsilon)^p \int_{0}^{\pi/2} e^{i \varphi p} \exp( -i\epsilon e^{-i\varphi} ) d \varphi,
	\end{equation}
	and, so,
	\begin{equation}
		\begin{aligned}
			\left | I_5 \right | &=  e^{-\pi y/2}\epsilon^x \left | \int_{0}^{\pi/2} e^{i \varphi p} \exp( -i\epsilon e^{-i\varphi} ) d \varphi \right | \\ &\le e^{-\pi y/2}\epsilon^x \int_{0}^{\pi/2} e^{-\varphi y}e^{-\epsilon \sin \varphi} d\varphi\\
			&\le \frac{\pi}{2} e^{-\pi y/2}\epsilon^x e^{\pi|y|/2},
		\end{aligned}
	\end{equation}
	which, as $x >0$, tends to zero when $\epsilon \to 0$.
\end{itemize}


\begin{thebibliography}{100}

\bibitem {lapena} L. de la Pena, A.M. Cetto and A. Valdes-Hernandez, \textit{The Emerging Quantum: The Physics Behind Quantum Mechanics} (Springer Publishing Company, New York, 2014).

\bibitem {kaled2} K. Dechoum, T.W. Marshall and E. Santos, Journal of Modern Optics  {\bf 47}, 1273-1287 (2000).
     
 \bibitem {kaled1} K. Dechoum and H.M. Franca, Found. Phys. {\bf 25}, 1599 (1995).

 \bibitem {milonni1} P.W. Milonni, \textit{The Quantum Vacuum, An Introduction to Quantum Electrodynamics} (Academic Press, Cambridge, 1994).
   
 \bibitem {boyer1} T.H. Boyer, Phys. Rev. {\bf 182}, 1374 (1969).

\bibitem {boyer2}  T.H. Boyer,  Phys. Rev. D {\bf 27}, 2906-2911 (1983).

\bibitem {boyer3} T.H. Boyer, Phys. Rev. D {\bf 29}, 1096-1098 (1984).

\bibitem {boyer4} T.H. Boyer, Phys. Rev. D {\bf 81}, 105024 (2010).
  
\bibitem {boyer5} T.H. Boyer, Phys. Rev. D {\bf 21}, 2137 (1980).
 
\bibitem {boyer6} T.H. Boyer, Phys. Rev. D {\bf 29}, 1089-1095 (1984).

 \bibitem {milonni2} P.M. Alsing, and P.W. Milonni, Am. J. Phys. {\bf  72},  1524-1529  (2004).

 \bibitem{planck} M.  Planck, Annalen der Physik {\bf309}, 553-563 (1901).

 \bibitem {milonni3} P.W. Milonni and M.L. Shih, Am. J. Phys. {\bf59}, 684(1991).
   
 \bibitem {cohen} J. Dalibard, J. Dupont-Roc and C. Cohen-Tannoudji,  J. Phys. France {\bf43}, 1617-1638 (1982).

 \bibitem {landsberg} P.T. Landsberg, \textit{Thermodynamics} (Intescience Publishers, Genebra, 1961), v. 2, p. 293.

 \bibitem{jaynes} E.T. Jaynes, Phys. Rev. {\bf 106}, 620 (1957).

 \bibitem {moriconi}  M. Moriconi, Eur. J. Phys.  {\bf  27}, 1409-1423 (2006).

 \bibitem {pathria} R.K. Pathria and P.D. Beale, \textit{Statistical Mechanics} (Elsevier, Oxford, 2011), p. 609.

 \bibitem {mandel}  L. Mandel and E. Wolf, \textit{Optical coherence and quantum optics} (Cambridge University Press, Cambridge, 1995).
   
 \bibitem {arfken}  G.B. Arfken, \textit{Mathematical Methods for Physicists} (Academic Press, Cambridge, 1985), 3º ed, p. 548.

 \bibitem {aitchinson} J.J.R. Aitchinson, Comtemp. Phys. {\bf 26}, 333 (1985).

 \bibitem {dice} L. de la Pe\~{n}a and A. M. Cetto, \textit{The Quantum Dice: An Introduction to Stochastic Electrodynamics}
   (Kluwer, The Netherlands, 1996). 
   
\end{thebibliography}
\end{document}